\documentclass[12pt]{article}

\usepackage{amsfonts}

\def\beq#1{\begin{equation}\label{#1}}
\def\eeq{\end{equation}}

\renewcommand{\theequation}{\arabic{section}.\arabic{equation}}
\catcode`\@=11 \@addtoreset{equation}{section}\catcode`\@=12

\newcommand{\bear}[1]{\begin{eqnarray}\label{#1}}
\newcommand{\ear}{\end{eqnarray}}
\def\ber#1{\begin{eqnarray}\label{#1} \nqq}
\def\eer{\end{eqnarray}}

\newcommand{\N}{{\mathbb N}}
\newcommand{\R}{{\mathbb R}}
\newcommand{\fnm}{\footnotemark}
\newcommand{\fnt}{\footnotetext}

\newcommand{\sh}{\sinh}
\newcommand{\ch}{\cosh}

 \def\nqq{\hspace*{-2em}}

\begin{document}

\begin{center}

\large\bf On generalized  black brane solutions in
the model with  multicomponent anisotropic fluid

\vspace{15pt}

\normalsize\bf

  \vspace{5pt}

\bf V.D. Ivashchuk\fnm[1]\fnt[1]{ivashchuk@mail.ru}

\vspace{5pt}

 \it  Center for Gravitation and Fundamental Metrology, \\
 Scientific Research Center of Applied Metrology ``Rostest'', \\ 
 Ozyornaya St. 46, Moscow 119361,  Russian Federation \\
 and \\
\it Institute of Gravitation and Cosmology, \\
Peoples' Friendship University of Russia (RUDN University),\\ 
Miklukho-Maklaya St. 6, Moscow 117198, Russian Federation \\

\end{center}

\vspace{15pt}

\begin{abstract}

A family of spherically $O(d_0 + 1)$-symmetric solutions in the model with
 $m$-component  anisotropic fluid is obtained.
The metrics  are defined on a manifold which contains a product of
 $n-1$ Ricci-flat ``internal'' spaces. The equation of state for
any $s$-th component is defined by a vector $U^s = (U^s_i)$
belonging to  $\R^{n + 1}$ and obeying inequalities 
$U^s_1 = q_s > 0$, $s = 1, \ldots,m$. 
The solutions  are governed by   moduli
functions $H_s$ which are solutions to (master) non-linear
differential equations with certain boundary conditions imposed.
It is shown that for coinciding $q_s = q$ there exists a subclass
of solutions with a horizon when $q = 1, 2, \ldots$ and
 $U^s$-vectors correspond to certain semisimple Lie algebras. An
extension of these solutions to block-orthogonal set of vectors
  $U^s$ with natural parameters $q_s$ coinciding inside blocks is
also proposed.  $q$-analogues of black brane/hole solutions
are presented, e.g. generalising $M_2 \cap M_5$ dyonic solution in $D
 =11$ supergravity and Myers-Perry  charged black hole solution
 in dimension  $D = 2 + d_0$.

\end{abstract}

\vspace{15pt}




\section{Introduction}

In  this paper we consider  spherically-symmetric solutions with a horizon in
the multidimensional model with 
multicomponent anisotropic fluid defined on warped product manifold 
$\R \times M_0 \times \ldots \times M_n$.

 Here we generalize solutions from
\cite{IMS}-\cite{I-bh-10} 
which may simulate  black brane
solutions appearing in the models with antisymmetric forms and
scalar fields. (For black brane solutions see
\cite{St}-\cite{IM-11} and references therein.)
Here we present a  new family of exact spherically-symmetric
solutions in the model with $m$-component anisotropic fluid  for
the following equations of state \cite{DI-03}
 \beq{0.1}
   p_r^s = - \rho^s /(2q_s-1), \qquad p_0^s = \rho^s/ (2q_s-1),
 \eeq
and
 \beq{0.2}
   p_i^s= \left(1-\frac{2U^s_i}{d_i} \right) \rho^s/(2q_s-1),
 \eeq
 $i > 1$, $s = 1, \ldots, m$, where for $s$-th component:
 $\rho^s$ is a density, $p_r^s$ is a radial pressure,
 $p_i^s$ is a pressure in $M_i$ ($i >1$)
  and   $q_s \neq 1/2$, $q_s > 0$.
 Parameters $U^s_i$ ($i > 1$)  and
    $q_s = U^s_1 > 0$
  obey certain relations described in Section 2.
 Here the manifold $M_0$ is
$d_0$-dimensional sphere ($d_0 \geq 2$) and $p_0^s$ is the pressure in $M_0$ 
(i.e. in one of  tangent directions). (For  multifactor cosmological models see
\cite{IMZ,IM5,LB,GIM} and refs. therein.)

 Special  solutions with  $q_s = 1$ were found in
 \cite{IMS} (for $m=1$), \cite{IMS2} (for orthogonal vectors $U^s = (U^s_i)$ and
 \cite{I-bh-10}. For $q_s > 0$   special solutions were
 obtained in \cite{DI-03} (for orthogonal $U^s$) and in
  \cite{DIM-02} (for $m=1)$. In  \cite{IM-11} the post-Newtonian
 parameters $\beta$ and $\gamma$ corresponding to the 4-dimensional
 section of the metric from \cite{I-bh-10} were calculated.

 Here we  show  that there exist  solutions with regular horizon when
 $U^s$-vectors correspond to simple roots of  certain semisimple Lie algebra
 and  $q_s$ are integer numbers obeying certain restrictions.
 (For simple Lie algebra $q_s$ should coincide: $q_s = q$ for all
 $s$.) We construct  generalized analogues  of black brane solutions.

The paper is organized as follows. In Section 2 the model with
multicomponent anisotropic  fluid is formulated.
In Section 3 a subclass of spherically symmetric solutions
(generalizing those from \cite{I-bh-10}) is presented. In Section
4 and a subclass of solutions with regular horizon corresponding
to integer $q_s$ is singled out. In Section 5 $q$-analogs of black
brane solutions are considered, e.g. a $q$-analogue of  $M2 \cap M
5$ dyonic solution. In Section 6  a $q$-analogue
of Myers-Perry (static) charged black hole solution \cite{MyersPerry} in dimension $D = 2 + d_0$
 is outlined. In Appendix the derivation of solutions is presented.

For the convenience of the reader, we add Table 1, where the main contributions of this paper and the author's previous papers on the topic are listed. In the table, the column ``derivation'' indicates (by $+$ or $-$) whether a complete derivation of the exact solution is presented in a cited paper. The solutions from \cite{IMS}-\cite{I-bh-10}
are special cases of the solutions presented in this article.

\begin{table}[h!]
      \centering
    \caption{Main contributions of this and  previous author's papers on black hole/brane solutions with $m$-component anisotropic fluid }
    \label{tab:my_table}
    \begin{tabular}{ccccc}
        \hline
        Reference &  $m$       & $q_s$        & $U^s$       & derivation \\
        \hline
         [1]  & $m =1$     & $q =1$       &  unique      &    -     \\
        \hline
        [2]  & $m \in \N$ & $q_s = 1$    & orthogonal   &    +     \\
         \hline
        [3] &  $m =1$     & $q \in \N$   &  unique       &    +     \\
          \hline
        [4] & $m \in \N$   & $q_s \in \N$ & orthogonal   &   +     \\
         \hline
         [5] & $m \in \N$  & $q_s =1$     & generic        &   -     \\
         \hline
      This paper & $m \in \N$ & $q_{(a)}\in \N$ & block-orthogonal  & +  \\
         \hline
    \end{tabular}
\end{table}

\section{The model}

Here, we consider a family of spherically symmetric   solutions to
 Hilbert-Einstein equations with an multicomponent anisotropic
 fluid matter source
  \beq{1.1}
   R^M_N - \frac{1}{2}\delta^M_N R = k T^M_N
  \eeq
defined on the manifold
  \beq{1.2}
     M = {\R}_{.} \times (M_{0}=S^{d_0}) \times (M_1 =
    {\R}) \times M_2 \times \ldots \times M_n
   \eeq
with the block-diagonal metrics
  \beq{1.2a}
      g=     e^{2\gamma (u)} du \otimes du  + \sum^{n}_{i=0} e^{2X^i(u)}   h^{[i]}.
   \eeq

Here $\R_{.} = (u_{-},u_{+})$ is an interval. The manifold $M_i$
with the metric $h^{[i]}$ is the Ricci-flat space of dimension
$d_{i}$, $i=1,2,\ldots,n$, and $h^{[0]}$ is standard metric on the
unit sphere $S^{d_0}$, $d_0 > 1$ (${\rm Ric}[h^{[0]}]=(d_0-1)h^{[0]}$),  
$u$ is radial variable, $\kappa$ is the multidimensional gravitational
constant, $d_1 = 1$ and $h^{[1]} = -dt \otimes dt$.

The  energy-momentum tensor is adopted in the following form
  \beq{1.5a}
    T^{M}_{N}= \sum_{s = 1}^{m} T^{(s)M}_{N},
  \eeq
where
 \beq{1.5b}
   T^{(s)M}_{N}= {\rm diag}( p^s_r,
    p^s_0 \delta^{m_{0}}_{k_{0}}, - \rho^s , p_2^s
   \delta^{m_{2}}_{k_{2}}, \ldots ,  p_n^s \delta^{m_{n}}_{k_{n}}).
 \eeq
The pressures $p_r^s = p_r^s(u) $,  $p_i^s = p_i^s(u)$  and the
densities $\rho^s = \rho^s(u)$  obey the relations (\ref{0.1}),
(\ref{0.2}).

As in \cite{IM5,GIM} we impose the ``conservation law''
restrictions
  \beq{1.con}
  \nabla_{M} T^{(s)M}_N = 0,
  \eeq
$s = 1, \dots, m$. In what follows we put $\kappa = 1$ for
simplicity. 

As it was mentioned in \cite{DI-03} it is more convenient for
finding of exact solutions, to write the stress-energy tensor in
cosmological-type form
 \beq{1.5} (T^{(s)M}_{N})= {\rm diag}(-{\hat{\rho^s}},{\hat p}_0^s
   \delta^{m_{0}}_{k_{0}}, {\hat p}_{1}^s
   \delta^{m_{1}}_{k_{1}},\ldots , {\hat p}_n^s
   \delta^{m_{n}}_{k_{n}}),
 \eeq
where $\hat{\rho}^s$ and  ${\hat p}_{i}^s$ are ``effective''
density and pressures of $s$-th component, respectively, depending
upon $u$. The physical density $\rho^s$ and pressures $p_i^s$ are
related to the ``effective''  ones by formulas
 \beq{1.7a}
   \rho^s = - {\hat p}_1^s,
    \quad p_r^s = - \hat{\rho}^s, \quad
     p_i^s = \hat{p}_i^s \quad (i \neq 1),
 \eeq
$s = 1, \dots , m$.

The equations of state may be written in the following form
 \beq{1.7}
    {\hat p}_i^s =\left(1-\frac{2U^s_i}{d_i}\right){\hat{\rho^s}},
 \eeq
where $U^s_i$ are constants, $s = 1, \ldots, m$; $i= 0,1, \ldots, n$
and $m \leq n$.

We impose the following restrictions on $U^s = (U^s_i) \in
\R^{n+1}$:

 \begin{eqnarray} \label{2.1}
  &&1^{o}.\quad U^s_0 = 0,
      \\ \label{2.1a}
     &&2^o.\quad U^s_1 = q_s >0, \quad  q_s \neq 1/2,
        \\ \label{2.1c}
  &&3^o. \quad ((U^s,U^l)) = (U^s_i G^{ij} U^l_j)
         \quad {\rm \ is \ non-degenerate \ matrix}.
         \end{eqnarray}

 Here
  \beq{2.2a}
     G^{ij}=\frac{\delta^{ij}}{d_i} + \frac{1}{2-D}
   \eeq
are components of the matrix inverse to the matrix of the
minisuperspace metric \cite{IMZ}
   \beq{2.2}
     (G_{ij}) = (d_i \delta_{ij} - d_i d_j),
   \eeq
  $i,j = 0, \ldots, n$,
and  $D=1+\sum\limits_{i=0}^n {d_i}$ is the total dimension.

In our case the scalar products (\ref{2.1c})
 are given by relations:
   \beq{2.1d}
    (U^s,U^l) =   \sum_{i=0}^{n} \frac{U^{s}_i U^l_i}{d_i}
     + \frac{1}{2-D} \left(\sum_{i= 0}^{n } U^s_i \right)
                      \left(\sum_{j= 0}^{n } U^l_j \right).
   \eeq
 The restrictions $1^{o}$  and  $2^{o}$ imply
     $K_s = (U^s,U^s)  > 0$  for all $s$ \cite{DI-03} (see Appendix). 
 In what follows we  will use a ``quasi-Cartan'' matrix

  \beq{2.1A}
  \quad (2(U^{s},U^{l})/ (U^{l},U^{l})) = (A_{sl}),
 \eeq
 which  is a non-degenerate matrix due to restrictions $3^{o}$  
  (or (\ref{2.1c})).
   This fact implies the restiction $m \leq n$.

\section{Exact solutions }

For the equations of state (\ref{0.1}) and (\ref{0.2}) with
parameters $U^s_i$ ($i > 0$) obeying (\ref{2.1c}) (see
(\ref{2.1d})) we obtain the following spherically symmetric
solutions  to the Hilbert-Einstein equations (\ref{1.1})

\begin{eqnarray} \label{12}
    g = J_{0} \left( F^{-1}  dR \otimes dR + R^{2} h^{[0 ]} \right) -
    J_1 F dt \otimes dt    + \sum_{i=2}^{n} J_{i} h^{[i ]},
 \\  \label{13}
    \rho^{s}= - (2q_s-1)\frac{A_s F^{q_s - 1}}{J_0 R^{2d_0}} \prod_{l=1}^m H_l^{-A_{sl}} ,
    \end{eqnarray}
 which may be derived by analogy with the black brane solutions
 \cite{IMp2,IMp3}. Here $F = 1 - 2 \mu R^{-d}$,  $d=d_0-1$,
  $h^{[0]}= d\Omega^2_{d_0}$ 
  is canonical metric on $S^{d_0}$ and ${\rm
 Ric}[h^{[j]}]$= 0 for $j > 0$,

\beq{2.3}
    J_{i} = \prod_{s =1}^m H_s^{-2 h_s U^{s i }},
\eeq
 $i = 0,1,...,n$;   $\mu >0$; 
   $A_s \neq 0$ are integration constants which appear 
   from solutions to equations (\ref{1.con}) (see Appendix)
  and
 \begin{eqnarray} \label{2.4}
    U^{s i} = G^{ij}U^{s}_{j}  = \frac{U^{s}_i}{d_i} + \frac{1}{2-D}
                \sum_{j=0}^{n}U^{s}_j,
  \\  \label{2.4a}
        h_s = K_s^{-1}, \quad  K_s = ({U^{s}},{U^{s}}).
\end{eqnarray} 
 
 It follows from  $1^{o}$ and (\ref{2.4}) that
   \beq{2.4b}
    U^{s 0} = \frac{1}{2-D} \sum_{j=1}^{n}U^{s}_j .
  \eeq

 Functions $H_s > 0$ obey the equations

 \beq{2.2.1}
  R^{d_0} \frac{d}{dR} \left[ R^{d_0} \frac{F }{H_s} \frac{d
  H_s}{dR} \right] =  B_s F^{q_s - 1} \prod_{l = 1}^{m}   H_{l}^{- A_{s l}},
 \eeq
 with $B_s = 2 K_s A_s$
 and the boundary conditions imposed:
   \beq{2.2.1a}
   H_s(R)  \to  H_s(R_0)= H_{s0} > 0  \quad
    {\rm for} \quad R \to R_0 \equiv (2 \mu)^{1/d},
   \eeq
   and
  \beq{2.2.1b}
  H_s (R = + \infty) = 1,
  \eeq
  $s = 1,..., m$.

The first boundary condition (\ref{2.2.1a})
   implies the fulfillment of the so-called  ``weak horizon condition'',
   which is an infinite propagation  time of  light as $R^{d} \to  2 \mu$.
   Indeed for radial null-geodesics we get 
   $ds^2 = J_{0}  F^{-1}  dR^2  -  J_1 F dt^2 = 0$,  which implies $t(R) \to + \infty$
   as $R \to R_0$, since $J_{0}(R) \to J_{0}(R_0) > 0$ and 
   $J_{1}(R) \to J_{1}(R_0) > 0$ as $R \to R_0$ due to (\ref{2.2.1a}).    
   For our metric obeying (\ref{2.2.1a}) this a necessary but not a sufficient condition 
   for the horizon at $R =  R_0$.
    In fact, an event horizon \cite{HE} at $R =  R_0$,  or equivalently (for our static 
    spherically-symmetric case) apparent one  \cite{HE},  appears for integrable 
    configurations with polynomial structure  of $H_s$ in terms of $2 \mu/R^{d} $ and natural  $q^s \in \N$, e.g., for  block-orthogonal     sets of vectors $U^s$, corresponding to semisimple Lie algebras (see Section 4 below). 
    The general proof of this fact (or conjecture)  will be the subject of a separate publication,
     though  it may be  verified for certain examples of solutions presented below (at least for 
     $m =1$).   The key point here is an analytical (or regular) behavior of functions $H_s(R)$ in the vicinity 
      of  $R_0$ for natural $q_s$, $s = 1,\dots, m$. 
      For the simplest solution with $D =4$, $m = d =1$ and $q \in \N$  see Ref. \cite{BolIvas}, 
    where  the global structure of the metric was analysed.

  The second one  (\ref{2.2.1b})  leads to  an asymptotically flatness
 of the  $(2 + d_0)$-dimensional section of the metric.

   Due to to (\ref{2.1}) and (\ref{2.4}) the metric may be
   rewritten as follows
   \begin{eqnarray} \label{12a}
    g = J_{0} \left[ F^{-1}  dR \otimes dR  + R^{2}  h^{[0]}  -
        \left(\prod_{s =1}^m H_s^{-2 q_s h_s } \right)
        F dt\otimes dt \right.
   \\ \nonumber \left.
         + \sum_{i=2}^{n} Y_{i} h^{[i ]} \right],
    \end{eqnarray}
  where
  \beq{2.3a}
  Y_i = \prod_{s =1}^m H_s^{-2 h_s U^{s}_i/d_i }.
  \eeq

 A complete derivation of the solution (\ref{12}), (\ref{13}) 
 is given in Appendix.

{\bf Remark.} The solution is valid for arbitrary Einstein space
($M_0,h^{[0]}$) obeying ${\rm Ric}[h^{[0]}]=(d_0-1)h^{[0]}$.

 A special   orthogonal case of this solution
 (when $(U^s,U^{l})= 0$ for  $s \neq l$)
 was considered earlier in \cite{DI-03} (for $m =1$ see
 \cite{DIM-02}). The solution with $q_s = 1$  was considered in
\cite{I-bh-10}; for special orthogonal case see also \cite{IMS2}
and \cite{IMS} (for $m=1$).

\section{Solutions with a  horizon}

Here we consider a subclass of black brane like solutions with a
(regular) horizon.  We  impose  the following (additional)
condition on the solutions
   \beq{2.2.1c}
     H_s(R) > 0  \ {\rm is \ analytic \ in} \ (R_{\epsilon}, +
     \infty),
   \eeq
  $s = 1,..., m$, where $R_{\epsilon}= (2 \mu)^{1/d} e^{-
  \epsilon}$ with some   $\epsilon > 0$.  Then  the metric (\ref{12}) has a 
  (regular) horizon  at $R^{d} =   2 \mu$.
   This is valid at least for known polynomial solutions to master equations for $H_s$ 
   and natural $ q_s$.

\subsection{One-block solutions}

Let us put
  \beq{2.12qs}
     q_s = q
  \eeq
for all $s$, where $q$ is a natural number:   $q= 1,2, \ldots$. In
this case there exists a subclass of solutions with  regular
horizon at $R^d = 2 \mu$.

Let us denote $z = R^{-d} \in (0, (2\mu)^{-1})$ and
      \beq{2.13}
      2  \mu Z =   1 - (1 - 2 \mu z)^{q},
      \eeq
where $Z  \in (0, (2\mu)^{-1})$ ($q > 0$).

We get a set of equations

\beq{5.3.1}
 \frac{d}{dZ} \left( \frac{(1 - 2\mu Z)}{H_s}
 \frac{d}{dZ} H_s \right) = \bar B_s
 \prod_{l =1}^{m}  H_{l}^{- A_{s l}}, \eeq
 where $H_s = H_s(Z) > 0$ for $Z  \in (0, (2\mu)^{-1})$
   and $\bar B_s =
  B_s/ (q d)^2 \neq 0$. Eqs. (\ref{2.2.1a}) and  (\ref{2.2.1b})
 read
 \begin{eqnarray} \label{5.3.2a}
  H_{s}((2\mu)^{-1} -0) = H_{s0} \in (0, + \infty), \\
 \label{5.3.2b} H_{s}(+ 0) = 1, 
 \end{eqnarray} 
  $s = 1,..., m$.

 The condition  (\ref{2.2.1c}) reads as follows
     \beq{5.3.2c}
     H_s(Z) > 0 \ {\rm is \ analytic \ in} \ (0, Z_{\epsilon}),
   \eeq
   $s = 1,..., m$, where $Z_{\epsilon}= (2 \mu)^{-1} e^{ \epsilon d}$
   with some $\epsilon > 0$.

   It was conjectured in \cite{IMp1}
 that  equations (\ref{5.3.1})-(\ref{5.3.2b})
 have  polynomial solutions  when  $(A_{s s'})$ is a  Cartan matrix for
 some  semi-simple finite-dimensional Lie algebra ${\bf G}$ of rank
 $m$.  In this case we get

 \beq{5.3.12}
 H_{s}(Z) = 1 + \sum_{k = 1}^{n_s} P_s^{(k)} Z^k, \eeq
 where $P_s^{(k)}$ are constants, $k = 1,\ldots, n_s$;
 $P_s^{(n_s)} \neq 0$, and

 \beq{5.2.20}
  n_s = b_s \equiv
            2 \sum_{l = 1}^m  A^{s l} ,
  \eeq
 $s = 1,..., m$, are the components of twice the  dual Weyl
 vector in the basis of simple  co-roots \cite{FS}.
 Here $(A^{sl}) = (A_{sl})^{-1}$.

 This conjecture  was verified for ${\bf A_m}$ and ${\bf C_{m+1}}$
 series of Lie algebras in \cite{IMp2,IMp3}. In the extremal case ($\mu
 = + 0$) an analogue of this conjecture was suggested
(implicitly) in \cite{LMMP}.

{\bf  ${\bf A_1} $ -case.} The simplest example occurs for $m=1$
 when $(A_{s l}) = (2)$ is a Cartan matrix for the simple Lie
algebra ${\bf A_1}$. In this case

 \beq{5.3.5}
 H_{1}(Z) = 1 + P_1 Z \eeq
 with $P_1 \neq 0$,  satisfying

 \beq{5.3.5a}
 P_1(P_1 + 2\mu) = -\bar B_1 = - 2 K_1 A_1/(qd)^2,
 \eeq
  and
  \beq{5.3.5b}
 \rho^{1}= - (2q-1)\frac{A_1 F^{q - 1}}{J_0 R^{2d_0}} H_1^{-2}.
  \eeq
  The physical condition $\rho^{1} > 0$ and $q = 1,2, \ldots $ imply
    $A_1 < 0$. In this case  there  exists a unique  number  $P_1 > 0$ obeying (\ref{5.3.5a})
 (since $K_1 > 0$) and  we are led to a solution from  \cite{DIM-02}.

{\bf  $A_2$-case.}
 For the Lie algebra $\cal G$ coinciding with  ${\bf A_2} = sl(3)$
  we get $n_1 = n_2 =2$ and

 \beq{5.4.1} H_{s} = 1 + P_s Z + P_s^{(2)} Z^{2},
 \eeq
 where $P_s=
 P_s^{(1)}$ and $P_s^{(2)} \neq 0$ are constants, $s = 1,2$.

 It was found in \cite{IMp1} that for $P_1 +P_2 + 4\mu \neq 0$
 (e.g. when all $P_s >0 $) the following relations take place

 \beq{5.4.5}
 P_s^{(2)} = \frac{ P_s P_{s +1} (P_s + 2 \mu )}{2
    (P_1 +P_2 + 4\mu)}, \qquad \bar B_s = - \frac{ P_s (P_s + 2 \mu
       )(P_s + 4 \mu )}{P_1 +P_2 + 4\mu}, \eeq
 $s = 1,2$.
 Here we denote $s+ 1 = 2, 1$ for $s = 1,2$, respectively.

   ``Master'' equations (\ref{5.3.1}) were integrated  in
  \cite{GrIvKim1,GrIvMel2} for Lie
  algebras ${\bf C_2}$ and ${\bf A_3}$, respectively.
  (For ${\bf D_4}$-solutions in the extremal case $\mu \to +0$ see \cite{LMMP}.)
  For general simple Lie algebras the solutions to master equations (\ref{5.3.1}) may be obtained by using
  so-called fluxbrane polynomials \cite{Ifbb}. (For fluxbrane polynomials see \cite{Iflux}, 
  and also \cite{Isym2017,BIsym2023}.)

\subsection{Block-orthogonal solutions}

Let  $S = \{1, \dots, m \}$ and

\beq{5.S}
 S=S_1 \cup\dots\cup S_k, \qquad  S_a \cap S_b =
\emptyset, \quad a \neq b, \eeq
$S_a \neq \emptyset$, i.e. the set
$S$ is a union of $k$ non-intersecting (non-empty) subsets
$S_1,\dots,S_k$. Here $S_1 < \dots < S_k$ and the ordering in
$S_i$ is inherited by the ordering in $S$.

 We put
 \beq{5.bo}
  (U^s,U^{l}) = 0,
  \eeq
  for all $s\in S_a$, $l \in S_b$, $a \neq b$;
 $a,b =1,\dots,k$.

 Let $q_s$  be coinciding inside blocks, i. e.
  \beq{5.12qs}
     q_s = q_{(a)},
   \eeq
for all $s \in S_a$, where   $q_{(a)}$ is a natural number:
 \beq{5.12q}
  q_{(a)}  = 1,2, \ldots
 \eeq
   for any $a =1,\dots,k$.

Master equations  (\ref{2.2.1}) in this case split into $k$ sets
of equations

\beq{5.3.1bo}
 \frac{d}{dZ_a} \left( \frac{(1 - 2\mu Z_a)}{H_s}
 \frac{d}{dZ_a} H_s \right) = \bar B_s
 \prod_{l \in S_a}  H_{l}^{- A_{s l}}, \eeq

 $s \in S_a$,  $\bar B_s =
  B_s/ (q_{(a)} d)^2 \neq 0$, and

      \beq{5.13a}
      2  \mu Z_a =   1 - (1 - 2 \mu z)^{q_{(a)}},
      \eeq
  $a =1,\dots,k$. 
  We remind that here $z = R^{-d}$.

Indeed, due to (\ref{5.S}) and (\ref{5.12qs}) we get 
\beq{5.12qqa}
     (q_s) = (\underbrace{q_{(1)},\dots,q_{(1)}}_{m_1},\ldots, \underbrace{q_{(k)},\dots,q_{(k)}}_{m_k}),
   \eeq
where $m_a = |S_a|$ is the number of elements in $S_a$ (corresponding to $a$-th block of vectors $(U^s, s \in S_a)$),
 $q_{(a)} \in \N$, $a =1,\dots,k$ and $m_1 + \ldots + m_k = m$. 
Due to block-orthogonal condition (\ref{5.bo}) the quasi-Cartan matrix (\ref{2.1A}) has a block-diagonal structure with
$k$ blocks,  i.e.
\beq{5.1A}
(A_{sl}) = {\rm diag}((A_{s_1 l_1}), \dots, (A_{s_k l_k})), 
\eeq
where $(A_{s_a l_a})$ is $m_a \times m_a$ matrix with indicies $s_a, l_a \in S_a$, $a =1,\dots,k$. 
Non-block-diagonal components of $(A_{sl})$ are zero, i.e. 
\beq{5.2A}
A_{s_a l_b} = 0, \qquad {\rm for} \quad a \neq b .
\eeq
Here $a,b = 1,\dots,k$. 

Due to relation (\ref{5.2A}) we obtain 
 $\prod_{l \in S}  H_{l}^{- A_{s l}} =  \prod_{l \in S_a}  H_{l}^{- A_{s l}}$ 
for all $s \in S_a$, where $a =1,\dots,k$. Owing to coincidence  of $q_s$ for $s \in S_a$,
see (\ref{5.12qs}), we can define a collective variable $Z_a$ 
from (\ref{5.13a}), which gives us master equations  (\ref{5.3.1bo})
for $a$-th block, where $a =1,\dots,k$.   
 
{\bf  ${\bf A_1} \oplus \ldots \oplus {\bf A_1}$ -case.} A simple
example occurs in the orthogonal case: $(U^s,U^{l})= 0$ for $s
\neq l$ \cite{BIM,IMJ}. In this case $(A_{s l}) = {\rm
diag}(2,\ldots,2)$ is a Cartan matrix for the semi-simple Lie
algebra ${\bf A_1} \oplus \ldots \oplus {\bf A_1}$ and

 \beq{5.3.5bs}
 H_{s} = 1 + P_s Z_s \eeq
 with $P_s \neq 0$,  satisfying

 \beq{5.3.5ab}
 P_s(P_s + 2\mu) = -\bar B_s = - 2 K_s A_s/(q_s d)^2,
 \eeq
 and
   \beq{5.3.5bb}
  \rho^{s}= - (2q_s -1)\frac{A_s F^{q_s - 1}}{J_0 R^{2d_0}} H_s^{-2},
   \eeq
   $s = 1,..., m$. 
   The physical condition $\rho^{s} > 0$ and $q_s \in \N $ imply
     $A_s < 0$ for all $s$. In this case  there  exist  unique set of  
     numbers  $P_s > 0$ obeying (\ref{5.3.5ab})
    (since all $K_s > 0$) and  we are led to a solution from  \cite{DI-03}.
 
 Analogously one can consider a case of semi-simple Lie
algebra which is a sum of $k$ simple Lie algebras: ${\bf G} = {\bf
G_1} \oplus \ldots \oplus {\bf G_k}$. In this case a natural
number $q_{(a)}$ should be assigned to $a$-th component ${\bf
G_a}$, $a =1,\dots,k$.

There exists also a special solution to eqs. (\ref{5.3.1bo})
 \cite{Br1,IMJ2,CIM}

 \beq{5.3.1boh}
  H_{s}(Z_a) = (1 + P_s Z_a)^{b_s}
  \eeq
 with $b_s$ from (\ref{5.2.20}) and parameters $P_s$
 coinciding inside $a$-th block , i.e. $P_s =
 P_{s'}$ for $s, s' \in S_a$.
 Parameters $P_s \neq 0 $ satisfy the relations

   \beq{5.3.1bob}
  P_s(P_s + 2\mu) = - \bar B_s/b_s,
   \eeq
where $b_s \neq 0$ and parameters $\bar B_s/b^s$  are also coinciding
inside blocks, i.e. $\bar B_s/b_s = \bar
 B_{s'}/b_{s'}$ for $s, s' \in S_a$. Here $a =1,\dots,k$.

\section{$q$-analogs of black brane solutions}

The solutions with a regular horizon from \cite{I-bh-10} allowed
us to simulate  certain intersecting black brane solutions
\cite{IMtop} in the model with antisymmetric forms (without scalar
fields)  when all $q_s =1$.

These solutions may be also generalized to the block-ortogonal
case (\ref{5.S})-(\ref{5.12q}) for a chosen set of  natural
$q_{(a)} \in \N$, $a = 1, ..., k$.

We put
   \beq{3.1-a}
    U^s_i  = q_{(a)} d_i , \quad i \in I_s;
    \qquad   U^s_i  = 0 , \quad  i \notin  I_s,
  \eeq
for all  $s \in S_a$, $a = 1, ..., k$.

Here  $I_s = \{ i_1, \ldots,  i_k \} \in \{1, \ldots n \}$ is the
index set \cite{IMtop} corresponding to brane submanifold
 $M_{i_1}  \times \ldots \times M_{i_k}$. All sets $I_s$ contain
  index $1$, corresponding to the time manifold $M_1 = \R$.

The relation   (\ref{3.1-a}) leads us to  the following relation
for  dimensions of intersections of brane submanifolds
(``worldvolumes'') \cite{IMJ,IMtop}:

  \beq{3.1b}
   q_{(a)} q_{(b)} \left[d(I_s \cap I_l) -\frac{d(I_s)d(I_l)}{D-2}\right] = (U^s,U^l), \eeq
  $s \in S_a$, $l \in S_b$; $a, b = 1, ..., k$.
 Here $d(I) \equiv  \sum_{i \in I} d_i$.
   For $a \neq b$  (due to (\ref{5.bo})) we obtain

      \beq{3.1c}
      d(I_s \cap I_l) = \frac{d(I_s)d(I_l)}{D-2},
      \eeq
 for all $s \in S_a$, $l \in S_b$, while
 for $a = b$ we get (see (\ref{2.1A}))
        \beq{3.1d}
      2 \left[d(I_s \cap I_l) - \frac{d(I_s)d(I_l)}{D-2}\right]
       = A_{sl} \left[d(I_l) - \frac{(d(I_l))^2}{D-2}\right],
      \eeq
 $s, l \in S_a$.

  Thus the intersection rules do not depend upon parameters
  $q_{(a)} > 0$. They depend upon the choice of
  quasi-Cartan matrix $A_{sl}$ (e.g. Cartan one), which has a
  block-diagonal structure due to  (\ref{5.bo}).

   $q$-analogues of   $M2 \cap M2$,  $M2 \cap M5$ and  $M5 \cap M5$
 solutions corresponding to the semisimple Lie algebra  ${\bf A_1} \oplus {\bf A_1}$
 were suggested in \cite{DI-03}.

 In \cite{I-bh-10} a  simulation of $M_2 \cap M_5$ dyonic configuration
  in $D = 11$ supergravity \cite{IMp1} was considered. It corresponds to the Lie
 algebra ${\bf A_2}$. Here we consider a $q$-extension of this
 solution with $q = 1, 2, \ldots$.

 The solution is  defined on the manifold

\beq{5.4.8} M =  (2\mu, +\infty )
 \times (M_0 = S^{2})  \times (M_1 = \R) \times M_{2} \times
 M_{3}, \eeq
 where ${\dim } M_2 =  2$ and ${\dim } M_3 =  5$.
 The $U^s$-vectors corresponding to fluid components
  obey  (\ref{3.1-a}) with  $I_1 = \{ 1, 2 \}$,
  $I_2 = \{ 1, 3 \}$ and $q_{(a)} = q$.

The solution reads as following
 \begin{eqnarray}
  g =  H_1^{1/(3q)} H_2^{2/(3q)} \biggl\{ F^{-1} dR \otimes
  dR + R^2   h[S^2]  - F  H_1^{-1/q} H_2^{-1/q}  dt\otimes dt
  \\ \nonumber
   + H_1^{-1/q} {h}^{[2]} + H_2^{-1/q} {h}^{[3]} \biggr\},
    \label{5.4.9} \\
     \rho^{1}= - \frac{(2q-1)A_1 F^{q -1}}{J_0 R^{4}}  H_1^{-2/q} H_2^{1/q},
     \\ \label{5.4.10a}
       \rho^{2}= - \frac{(2q-1)A_2 F^{q -1}}{J_0 R^{4}}  H_1^{1/q} H_2^{-2/q},
    \label{5.4.10b}
 \end{eqnarray}

 where $F = 1 - 2\mu / R$, $J_0 = H_1^{1/(3q)} H_2^{2/(3q)}$; $h[S^2]$ is the canonical
 metric on 2-dimensional sphere $S^2$, $h^{[2]}$ and  $h^{[3]}$ are
 Ricci-flat metrics of Euclidean signatures defined on the manifolds
 $M_2$ and $M_3$, respectively; $\mu > 0$ and $H_s(Z)$
 are defined in (\ref{5.4.1}), with $Z = Z(z,q)$ from   (\ref{2.13}),
     $z = R^{- 1}$ and parameters
 $P_s$, $P_s^{(2)}$, $\bar B_s  = 4 A_s$,  $s =1,2$,
 obey  (\ref{5.4.5}). Here $K_s = (U^s,U^s)=2 q^2$  for  $s
=1,2$.

This  solution for $q = 1$ simulates ${\bf A_2}$-dyon from
\cite{IMp1} consisting of an electric $M2$-brane with a
worldvolume isomorphic to $(M_1 = \R) \times M_{2}$ and a magnetic
$M5$-brane with a  worldvolume isomorphic to $(M_1 = \R) \times
M_{3}$ \cite{I-bh-10}. The branes are intersecting on the time
manifold $M_1 = \R$.

{\bf Hawking temperature.} 
The Hawking temperature corresponding to the solution with a horizon
(\ref{12}) has the following form (see \cite{Y})

\beq{6.1}
 T_H =   \frac{d}{4 \pi (2 \mu)^{1/d}}
  \prod_{s = 1}^m H_{s0}^{- q_s h_s}, \eeq
where $H_{s0}$ are defined in  (\ref{2.2.1a}).

For the $q$-extension of dyonic solution from the previous
subsection we get

  \beq{6.2}
 T_H =   \frac{1}{8 \pi  \mu}
   (H_{10}H_{20})^{- 1/(2q)}. \eeq

Here
 \beq{6.2h} H_{s0} = 1 + P_s (2 \mu)^{-1} + P_s^{(2)} (2 \mu)^{-2},
 \eeq
where $\mu > 0$ and parameters $P_s > 0$,  $P_s^{(2)} > 0$ obey
(\ref{5.4.5}) with  $\bar B_s  = 4 A_s$,   $A_s < 0$.

For fixed parameters $\mu > 0$ and $P_s > 0$ (and hence $A_s < 0$)
 the values $H_{s0} > 1$ do not depend upon $q$. The  Hawking temperature $T_H (q)$
is monotonically increasing in a sequence $q = 1, 2, \dots$ and
tends to the Schwarzschild value $T_H (+ \infty) =   1/(8 \pi
\mu)$ as $q \to + \infty$. In this limit the metric (\ref{5.4.9})
tends to
 $g_{Sch} +  h^{[2]} + h^{[3]}$, where  $g_{Sch}$ is the Schwarzschild
 metric.  The   densities $\rho^{s}$ and pressures $p_i^s$
 vanish as $q \to + \infty$.

   For fixed parameters $P_s > 0$ and $q$ the Hawking temperature
   has the following asymptotical  behaviour
    $T_H (\mu) \sim {\rm const} \times \mu^{(2/q) -1}$
   as $\mu \to + 0$. It is divergent for $q > 2$, i. e.
   $T_H (+ 0) = + \infty$. For $q =1$ we get $T_H (+ 0) = 0$
   while for $q =2$ we find $T_H (+ 0) = T_0 > 0$.

 We note that the analysis  of $T_H$  may be of relevance  for possible search
 of  dual holographic models (at least for $q=1$ \cite{I-bh-10}) in a context of (an some version of) $AdS/CFT$
 approach \cite{Mald,GubKlPol,Witten}. 
 
\section{ $q$-analogue of  Myers-Perry  solution}

Here we consider a special spherically-symmetric black hole solution in dimension 
$D = 2 + d_0$  for the case $m = n = 1$, i.e. when we deal 
with one-component fluid and the ``internal'' space is absent.   

So, we explore the case of the manifold with metric of 
signature $(+, \dots,+,-)$
\begin{equation}
\label{N.1}
  M = R_{(\rm radial)}\times S^{d_0}\times R_{(\rm time)},
\end{equation}
where $S^{d_0}$ is $d_0$-dimensional sphere.

The energy-momentum tensor of anisotropic fluid is taken as
\begin{equation}
\label{N.2}
 (T^{\mu}_{\nu})={\rm diag}\left(p_r,\ p_0, \dots, p_0, \ -\rho  \right),
\end{equation}
and the equations of state read
\begin{equation}
\label{N.3}
  p_r = -\rho  (2q-1)^{-1}, \qquad p_0 = - p_r.
\end{equation}
Here $\rho$ is the mass density, $p_r$  and $p_0$  are  pressures in radial and orthogonal 
(to radial) directions, respectively.   

\newpage

The solution has the following form:
\begin{eqnarray} 
\label{N.4}
  ds^2= g_{\mu \nu} dx^{\mu} dx^{\nu}  = (H(R))^{2/(qd)}\left[
 \frac{dR^2}{1-\frac{2\mu}{R^d}}
 + R^2 d\Omega^2_{d_0} \right.
 \\ \nonumber 
\left. -(H(R))^{-2(d+1)/(qd)}\left(1-\frac{2\mu}{R^d}\right) dt^2\right],
\end{eqnarray}
\begin{equation}
 \label{N.5}
 \kappa \rho  = \frac{d(d+1)(2q-1)P(P+2\mu)(1-2\mu R^{-d})^{q-1}}{2(H(R))^{2+\frac{2}{qd}}\; R^{2d_0}},
\end{equation}
where the function $H(R)$ reads as follows:
\begin{equation}
 \label{N.6}
 H(R) = 1+\frac{P}{2\mu}
 \left[1-\left(1-\frac{2\mu}{R^d}\right)^q \right].
\end{equation}
The metric on the sphere $S^{d_0}$ is denoted by
 $d\Omega^2_{d_0}$, $d_0 = d + 1$; parameters  $P > 0$, $\mu > 0$ are arbitrary. 
 
 The solution (\ref{N.4})-(\ref{N.6}) just follows from general formulas for $m = n = 1$
 when we put for components of $U^1$-vector 
 \begin{equation}
 \label{N.6U}
     U^1_0 = 0, \qquad U^1_1 = q.
 \end{equation}
 Indeed, by using relation  (\ref{2.1d}) we obtain 
 \begin{equation}
 \label{N.6Kh}
     K_1 = (U^1,U^1) = \frac{q^2 d}{d+1}, \qquad h_1 =  K_1^{-1} = \frac{d+1}{q^2 d}, 
  \end{equation}
 which implies the relation for the metric (\ref{N.4}) (see (\ref{12})).
 For parameters $A = A_1$ and $P = P_1$ we find from (\ref{5.3.5a}) and (\ref{N.6Kh})
  \begin{equation}
 \label{N.6PA}
      P(P+2\mu) =  - \frac{2A}{d(d+1)}, 
   \end{equation}
 which implies the relation for the density (\ref{N.5}) (see (\ref{13})). For $q = 1$
 the metric (\ref{N.4}) is coinciding with the metric of the spherically-symmetric black hole
 in dimension $d_0 + 2$ from \cite{MyersPerry}.
 
 Originally, we put 
  $R^d > 2\mu$
 but the domain of definition of the metric may be extended (for $d = 1$ and $D=4$ see   \cite{BolIvas} ). 
We note that (\ref{N.6}) implies 
\begin{equation}
 \label{N.7}
 H(R) = 1+\frac{q P}{R^d} + o \left(\frac{2 \mu}{R^d} \right),
\end{equation}
as $R \to + \infty$. Due to this relation we obtain the 
asymptotically flatness of the metric (\ref{N.4}) 
since it tends to the metric
   $ds^2_{as} = dR^2 + R^2 d\Omega^2_{d_0} - dt^2$,
as $R \to + \infty$, which is a flat one.

It should be  also 
noted that the in the $D = 4$  case  this solution (with $d_0 = 2$) was
 presented earlier in  \cite{DIM-02}, while exploration of qusinormal modes and
null geodesics was considered recently in \cite{BolIvas} and  \cite{IvBolBKMNZ},
respectively. 


\section{Conclusion}
In this paper we have considered a family of spherically symmetric
solutions in the model with  $m$-component  anisotropic fluid when
the equations of state (\ref{0.1})- (\ref{0.2}) (with certain
restrictions on $U^s_i$-parameters) were imposed. The metric of
any solution contains $(n -1)$ Ricci-flat ``internal'' space
metrics and depends upon a set of parameters $U^s_i$ which define
the the equations of state of fluid components.  The solutions
under consideration are governed by   moduli functions $H_s$
obeying non-linear differential  equations with certain boundary
conditions imposed.

The parameters $U^s_1 = q_s > 0$  play a special role among all
other parameters $U^s_i$.
 Here we have shown  that for coinciding $q_s =
q$ there exists a subclass of solutions with regular horizon when
$q = 1, 2, \ldots$ and $U^s$-vectors correspond to simple roots of
certain semisimple Lie algebra. The regularity of horizon follow
from the polynomial structure of the functions $H_s(Z)$ and  $Z =
Z(z)$ (see (\ref{2.13})).

We have also generalized one-block  solutions with regular horizon
to the case of  block-orthogonal set of vectors $U^s$. In this
case several natural parameters $q_{(a)}$ occur, $a = 1, \dots,
k$, where $k$ is a number of blocks. We have also suggested
generalized analogues of black brane solutions in the
 block-orthogonal case.
 
  We have  shown that for a
 fixed  parameter of extremality $\mu > 0$ and fluid parameters
 $P_s > 0$ the Hawking temperature $T_H$ is monotonically
 increasing in a sequence $q = 1, 2, \dots$ and tends to the
 Schwarzschild value $1/(8 \pi \mu)$ as $q \to + \infty$.
 
As examples we have presented  a $q$-analogue of $M_2 \cap M_5$
dyonic solution in $D =11$ supergravity (corresponding to Lie
algebra $A_2$)  and $q$-analogue of $(2+d_0)$-dimensional Myers-Perry 
static charged black hole solution   ($q = 1, 2, \ldots$). 

\newpage 

\vskip6pt

{\bf Acknowledgement}

The author is grateful to  A.P. Yefremov and  S. Cotsakis for encouraging in submitting this paper.
The author also  thanks the anonymous referee for the insightful comments, which have improved the presentation and consistency of the manuscript.

\renewcommand{\theequation}{\Alph{subsection}.\arabic{equation}}
\renewcommand{\thesection}{}
\renewcommand{\thesubsection}{\Alph{subsection}}
\setcounter{section}{0}

\section{Appendix}

Here we present the derivation of the main solution from Section 3.

\subsection{Toda-type Lagrangian}

Here we deal with Hilbert-Einstein equations (\ref{1.1})  with an multicomponent anisotropic
 fluid matter source  defined on the manifold  (\ref{1.2}) with the  metric 
  \beq{A1.2a}
       g=    w e^{2\gamma (u)} du \otimes du  + \sum^{n}_{i=0} e^{2X^i(u)}   h^{[i]}.
    \eeq
 Here we slightly extend the metric  (\ref{1.2a}) by considering the Ricci-flat space 
 $(M_1,  h^{[1]})$ of arbitraty dimension $d_1$ and signature  
 and including the sign factor $w = \pm 1$, which gives
 us the spherically symmetric static metric for $w = 1$, $ h^{[1]} = - dt \otimes dt$
 and cosmological metric for $w = - 1$.

Here we use the stress-energy tensor of  $s$-component written in  ``cosmological-type'' form. 
The ``conservation law'' equations   $\nabla_{M} T^{(s) M}_N = 0$
may be written, due to relations (\ref{1.2a}) and (\ref{1.5}) in
the following form:
   \beq{A5.7}
   \dot{\hat{\rho^s}} +\sum_{i=0}^n
     d_i\dot{X^i}({\hat{\rho^s}} +{\hat p}_i^s )=0,
   \eeq
$s = 1, \dots, m$.
Using the equation of state (\ref{1.7}) we get
  \beq{A5.7a}
   \hat{\rho}^s = - w A_s e^{2U^s_i X^i -2 \gamma_0 },
  \eeq
$s = 1, \dots, m$,
where $\gamma_0(X)= \sum\limits_{i=0}^{n} d_{i}X^{i}$ and $A_s$
are constants.

The Einstein equations (\ref{1.1})  with the relations (\ref{1.7})
and (\ref{A5.7a}) imposed are equivalent to the Lagrange equations
for the Lagrangian
  \beq{lag} L = \frac{1}{2}
   e^{-\gamma+\gamma_0(X)}G_{ij}\dot{X}^{i}\dot{X}^{j}
     -e^{\gamma-\gamma_0(X)} V,
  \eeq
where
   \beq{A5.32n}
    V=  \frac{1}{2} w d_0 (d_0 -1) \exp(2U^0_i X^i) +
    \sum_{s= 1}^{m} A_s \exp(2 U^s_i X^i)
   \eeq
is the potential and the components of the minisupermetric
$G_{ij}$ are defined in (\ref{2.2}),
  \beq{A5.8}
    U^{0}_i X^i = -X^0 + \gamma_0(X), \qquad U^{0}_i  =- \delta^0_i + d_i,
  \eeq
 $i = 0, \ldots, n$. (For cosmological case $w = -1$  see \cite{IM5,GIM}).

For $\gamma=\gamma_0(X)$, i.e. when the harmonic time gauge is
considered, we get the set of Lagrange equations for the
Lagrangian
  \beq{A5.31n}
    L=\frac12  G_{ij} \dot X^i \dot X^j-V
  \eeq
with the zero-energy constraint imposed
  \beq{A5.33n}
   E=\frac12  G_{ij} \dot X^i \dot X^j + V =0.
  \eeq

It follows from the restriction
   $U_0^s = 0$ that
  \beq{A5.43a}
    (U^0,U^s)  \equiv U^0_i G^{ij}U_j^s = 0.
  \eeq

Indeed, the contravariant components $U^{0i}= G^{ij} U^0_j$ 
(for $G^{ij}$ see (\ref{2.2a}) ) are the following ones 

\beq{5.43b} U^{0i}=-\frac{\delta_0^i}{d_0}. \eeq

Then we get $(U^0,U^s)  = U^{0i} U_i^s = - U_0^s/d_0 =0$. In what
follows we also use the formula
\beq{A5.43c} 
 K_0 =(U^0,U^0)   = \frac{1}{d_0} - 1  < 0 
\eeq for $d_0 >1$.

Using (\ref{2.1}) and (\ref{2.1a}) one can be readily  prove that
\beq{A5.43s} 
K_s =(U^s,U^s) > 0 
\eeq
for all $s >0$. Indeed, minisupermetric (and scalar product $( ., .)$) has the signature 
$(-,+,\ldots,+)$ \cite{IMZ}, vector $U^0$ is time-like and orthogonal to any vector $U^s \neq
0$. Hence all  vectors $U^s$ belong to a subspace $V_{ort} \subset \R^{n+1} $ of vectors orthogonal to $U^0$.
The scalar product $( ., .)$ restricted on $V_{ort} $ is positive definite and hence 
$(U^s,U^s) \geq 0$ for all $s >0$. Relation $(U^s,U^s) = 0$ would only take place when $U^s = (U^s_i = 0)$. 
But  all vectors $U^s$ are non-zero due to condition (\ref{2.1a}). 
Thus, the relations (\ref{A5.43s}) are proved.

\addtocounter{section}{1} \setcounter{equation}{0}

\subsection{Cosmological type solutions}

Exact solutions to field equations have the following form 
(see \cite{IK} for special brane vectors $U^s$ ) 
  \beq{A5.34n}
    X^i(u)= - \sum_{s=0}^{m} \frac{U^{s i}}{(U^s,U^s)}
     \ln f_s(u)  + c^i u + \bar{c}^i,
  \eeq
$i = 0, \dots, n$, where   $c^i, \bar{c}^i$ are integration constants; 
and vectors $c=(c^i)$ and $\bar c=(\bar c^i)$ are dually-orthogonal to co-vectors
 $U^{s }=(U^{s}_i)$,
i.e. they satisfy the linear constraint relations
  \begin{eqnarray}
   \label{A5.47n}
   U^0(c)= U^0_i c^i = -c^0+\sum_{j=0}^n d_j c^j=0, \\
   \label{A5.48n} U^0(\bar c)= U^0_i \bar c^i =
   -\bar c^0+\sum_{j=0}^n d_j \bar c^j=0, \\
   \label{A5.49n} U^s(c)= U^s_i c^i=0,\\
   \label{A5.50n} U^s(\bar c)=  U^s_i \bar c^i=0,
  \end{eqnarray}
     $s = 1, \dots, m$.

Here 
\beq{B1.19}
      f_a = \exp( - q^a),
\eeq
$a = 0,1, \dots, m$,
where $(q^a) = (q^a(u))$ are  solutions to Toda-type equations
 \begin{eqnarray} 
 \label{B1.20a}
\ddot{q^0} = -  B_0 \exp( 2 q^{0} ),
\\
\label{B1.20}
\ddot{q^s} = -  B_s \exp( \sum_{l = 1}^m A_{s l} q^{l} ),
 \end{eqnarray}
with
\beq{B1.21}
   B_0 = 2 K_0 A_0 =  (-w)(d_0 -1)^2, \qquad    B_s = 2 K_s A_s,  
\eeq
$s = 1, \dots, m$. Here, as in \cite{IK}, the equations of motion are 
splitted into two independent set of equations 
(\ref{B1.20a}) and (\ref{B1.20}) due to relation (\ref{A5.43a}).

The first equation  (\ref{B1.20a}) is just a Liouville
equation while equations (\ref{B1.20}) are Toda-like 
equations governed by non-degenerate (``quasi-Cartan'') matrix $(A_{sl})$ from  (\ref{2.1A}).

The Liouville equation  
 (\ref{B1.20a}) corresponds to Lagrangian
\beq{B1.3aL}
L_{0} = \frac{1}{2} h_0  (\dot{q^0})^2 -   A_0  \exp( 2 q^{0} ),
\eeq
where $h_0 = K_0^{-1}$ and has the (energy) integral of motion
\beq{B1.3aE}
E_{0} = \frac{1}{2} h_0  (\dot{q^0})^2 +   A_0  \exp( 2 q^{0} ).
\eeq
 
Toda-type equations (\ref{B1.20}) are governed by the
 Lagrangian
\beq{B1.3L}
L_{TL} = \frac{1}{4}  \sum_{s,l = 1}^m
h_s  A_{s l} \dot{q^s}\dot{q^{l}}
-  \sum_{s = 1}^m A_s  \exp( \sum_{l = 1}^m A_{s l} q^{l} ),
\eeq
where $h_s = K_s^{-1}$ and has the  (energy) integral of motion
\beq{B1.3E}
E_{TL} = \frac{1}{4}  \sum_{s,l = 1}^m
h_s  A_{s l} \dot{q^s}\dot{q^{l}}
+  \sum_{s = 1}^m A_s  \exp( \sum_{l = 1}^m A_{s l} q^{l} ).
\eeq


The zero-energy constraint, corresponding to the solution
(\ref{A5.34n}) reads
  \beq{A.17}
    E = E_0 + E_{TL}   + \frac{1}{2} G_{ij}c^ic^j=0 . 
  \eeq

The solution to Liouville equation (\ref{B1.20a}) gives us
\begin{eqnarray} \label{B1.23}
f_0(u) =R \sh(\sqrt{C_0}u), \ C_0 >0, \ w = 1,
\\ \label{B1.24}
R \sin(\sqrt{|C_0|}u), \ C_0 < 0, \ w = 1,   \\ \label{B1.25}
R \ch(\sqrt{C_0}u),  \ C_0 > 0, \ w = - 1,   \\ \label{B1.26}
 (d_0 -1) u, \ C_0 =0, \ w = 1,
\end{eqnarray}
where  $R =  (d_0 - 1)/|C_0|^{1/2}$ and $C_0$ 
is a constant obeying 
\beq{B1.3aCE}
C_0 \frac{d_0}{d_0-1} = - 2 E_0.
\eeq
(Generally, the variable $u$ in the above relations for $f_0(u)$ should be replaced with 
$\varepsilon(u - u_0)$, where $u_0$ is an arbitrary constant and $\varepsilon = \pm 1$. 
Without loss of generality, we set $u_0 = 0$ and $\varepsilon = 1$.)

Due to relations (\ref{A5.47n}) and (\ref{B1.3aCE})
zero-energy constraint (\ref{A.17}) may be rewritten as
\beq{B1.30a}
 C_0 \frac{d_0}{d_0-1}= 2 E_{TL} +
 \sum_{i=1}^nd_i(c^i)^2+
 \frac1{d_0-1}\left(\sum_{i=1}^nd_ic^i\right)^2.
\eeq

Let us consider a family of one-variable sector
solutions to field equations corresponding to the action
(\ref{1.1}) and depending upon one variable $u$
(for brane vectors $U^s$ see  \cite{IK}).
The solutions read 
\begin{eqnarray}
g= [f_0(u)]^{2d_0/(1-d_0)}\exp(2c^0 u + 2 \bar c^0) \times
\nonumber  
\\ 
\times \Bigl[ w \biggl(\prod_{s \in S} [f_s(u)]^{2 a_s } \biggr)  du \otimes du 
+ \biggl(\prod_{s \in S} [f_s(u)]^{2 a_s^0 } \biggr)  f_0^2(u) h^{[0]}  \Bigr]
\nonumber  \\
 + \sum_{i = 1}^{n} \Bigl(\prod_{s\in S}
[f_s(u)]^{ 2 a_s^i } \Bigr)
\exp(2c^i u+ 2 \bar c^i) h^{[i]}, \label{5.g}
\end{eqnarray}
where 
\begin{eqnarray}
a^i_s = - U^{si}/(U^s,U^s) = - h_s U^{si} \label{5.ais},  \\ 
a_s = a^i_s d_i = - (U^{\Lambda},U^s)/(U^s,U^s) = 
- h_s \sum_{i=0}^{n} d_i U^{si}, \label{5.as}
\end{eqnarray}
$i =0, \dots , n$; $s \in S$,  where vector $(U^{\Lambda}_i) = (d_i)$  corresponds to
the $\Lambda$-term.

For the ``effective'' density of $s$-component we get
\begin{equation} \label{5.r}
 \hat{\rho}^s = - w A_s |f_0|^{2d_0 /(d_0 - 1)} 
 e^{- 2 c^0 u  - 2 \bar{c}^0} 
  \prod_{s' \in S} f_{s'}^{ - A_{s s'} - 2 a_{s'}}, 
  \end{equation}
 $s \in S$.

\subsection{ Solutions with a horizon}

Here  we 
consider the static spherically symmetric metric with $w = 1$.
We also  assume that
\beq{2.2b}
M_1 = \R, \qquad h^{[1]} = - dt \otimes dt,
\eeq
i.e.  $M_1$ is a time manifold.
 
Here we will use  relations (\ref{A5.43s}) and 
condition (\ref{2.1c}) 
\begin{equation} 
 {\rm det}((U^s,U^l)) \neq 0. \label{B1.18b}
\end{equation}
Here and in what follows we denote 
$S = \{1, \dots, m \}$. Relation (\ref{B1.18b})
implies that all vectors $U^s$ (belonging to $V_{ort}$) are linearly independent and hence 
$m \leq n = {\rm dim} V_{ort} $. Moreover, the symmetric matrix  $((U^s,U^l))$ is positive definite,
since $(U^s,U^l) = <U^s,U^l>$, where $< ., .>$ is positive-definite scalar product on 
subspace $V_{ort}$, obtained by restriction of $( ., .)$ on $V_{ort} \times V_{ort}$.

In this case relation (\ref{B1.23}) reads
\begin{eqnarray} \label{1.23s}
f_0(u) =  d \frac{\sh(\sqrt{C_0} u)}{\sqrt{C_0}},
\ C_0 > 0,
\\ \label{1.26s}
  ud, \ C_0=0,
\end{eqnarray}
where
$ d = d_0 -1$.

Now, we put $C_0 > 0$. 
Let us consider the null-geodesic equations
for the light ``moving'' in the radial direction
(following from $ds^2 =0$):
\begin{eqnarray} \label{C2.4}
 \frac{dt}{du} = \pm \Phi, \\ \label{C2.4a}
 \Phi =  f_0^{d_0/(1-d_0)} e^{(c^0 - c^1) u +  \bar c^0 - \bar c^1}
 \prod_{s \in S} f_s^{h_s  U_1^s},
\end{eqnarray}
equivalent to
\beq{C2.5}
 t - t_0 = \pm \int_{u_0}^{u} d \bar u  \Phi(\bar u),
\eeq
where $t_0, u_0$ are constants.

Let us consider   solutions
(defined on some interval $[u_0, +\infty)$) 
which obey a ``weak  horizon
condition'' at $u = + \infty$ satisfying
\beq{C2.6}
  \int_{u_0}^{ + \infty} d  u  \Phi( u) = + \infty.
\eeq

Here we restrict ourselves to  solutions with
$C_0 > 0$ and linear asymptotics at infinity
\beq{C2.7}
q^s = - \beta^s u + \bar \beta^s  + o(1),
\eeq
$u \to +\infty$, where $\beta^s, \bar \beta^s$ are
constants, $s \in S$. This relation gives us an
asymptotical solution to  Toda type eqs. (\ref{B1.20}) if
\beq{C2.8}
\sum_{s' \in S} A_{s s'} \beta^{s'} > 0
\eeq
for all $s \in S$. In this case the energy  (\ref{B1.3E})
reads
\beq{C2.9}
E_{TL} = \frac{1}{4}  \sum_{s,s' \in S}
h_s A_{s s'}  \beta^s \beta^{s'}.
\eeq

For  the function  (\ref{C2.4a}) we get
\beq{C2.10}
\Phi(u) \sim \Phi_0 e^{\beta u}, \quad u \to +\infty,
\eeq
where  $\Phi_0 \neq 0$ is  constant,
\beq{C2.11}
 \beta = c^0 - c^1 + \sqrt{C_0} h_0
+ \sum_{s \in S} \beta_s h_s  ,
\eeq
and
  $h_0 = (U^0, U^0)^{-1} = \frac{d_0}{1 - d_0}$.
Horizon at $u = + \infty$   take place if 
\beq{C2.12a}
  \beta \geq 0.
\eeq
Let us introduce dimensionless parameters
\beq{C2.13}
b^s = \beta^s / \sqrt{C_0}, \qquad
b^i =  c^i / \sqrt{C_0},
\eeq
where $s \in S$, $i = 0,1, \dots, n$; $C_0 > 0$.

Thus, a horizon at $u = + \infty$
corresponds to a point $b = (b^s,b^i) \in  \R^{m + n + 1}$
satisfying the relations following from
(\ref{A5.47n}), (\ref{A5.48n}), (\ref{A5.49n}),(\ref{A5.50n}),(\ref{C2.8}), (\ref{C2.9})
and (\ref{C2.11})-(\ref{C2.13}):
\begin{eqnarray} \label{C2.14}
U^r_i b^i= 0,   \qquad r = s, 0; \ s \in S, \\ \label{C2.15}
\frac{1}{2}  \sum_{s,s' \in S} h_s A_{s s'}  b^s b^{s'}
+  G_{ij} b^i b^j = |h_0|, \\
\label{C2.16}
\sum_{s' \in S} A_{s s'} b^{s'} > 0,  \\ \label{C2.17}
 f(b) \equiv
 b^0 - b^1  + \sum_{s\in S} b_s h_s  U^s_{1}  \geq |h_0|.
\end{eqnarray}

{\bf Proposition.} {\em  The point $b = (b^s,b^i)$ satisfying
relations (\ref{C2.14})-(\ref{C2.17}) exists only if
\beq{2.18}
U^s_{1} > 0, \quad \forall s \in S,
\eeq
and is unique: $b = b_0$, where
\begin{eqnarray} \label{C2.19}
b_0^i  = - \delta^{i}_{1} + h_0 U^{0 i}  +
\sum_{s\in S}  h_s b_0^s U^{s i},  \\ \label{C2.20}
b_0^s = 2 \sum_{s' \in S} A^{s s'} U_1^{s'},
\end{eqnarray}
where $s \in S$, $i = 0, \dots, n$,
and the matrix $(A^{s s'})$ is inverse to the matrix
$(A_{s s'}) = ( 2 (U^s, U^{s'})/ (U^{s'}, U^{s'}))$
($s, s' \in S$). }

{\bf Proof.} Let ${\cal E}$ be a manifold described
by relations (\ref{C2.14})-(\ref{C2.15}). This manifold is an ellipsoid.
Indeed, due to the fact that
the matrix $(G_{ij})$ has a signature $(-,+, \ldots,+)$
 \cite{IMZ},
$(U^0,U^0) < 0$,  $(U^0,U^s) = 0$, $K_s =(U^s,U^s) >  0$ for all $s \in S$,
 and (\ref{B1.18b}), the symmetric matrices
 $(B_{s s'}) = (U^s, U^{s'})$ and $(h_s A_{s s'}) = (h_s B_{s s'} h_{s'} )$
  ($h_s = K_s^{-1} $) are positive
definite ones and all $h_s > 0$, $s \in S$. Then, the quadratic form
in (\ref{C2.15}) has a pseudo-Euclidean signature.
Due to $(U^0,U^0) < 0$ the intersection of the
hyperboloid (\ref{C2.15}) with the (multidimensional)
plane $U^0_i z^i = 0$ gives us an ellipsoid. 
Its intersection with the planes $U^s_i z^i = 0$, $s \in S$, leads us to a (final) ellipsoid,
 coinciding with ${\cal E}$, which is a compact submanifold. 

Let us consider  the function $f_{|} : {\cal E} \rightarrow \R$
that is a restriction of the linear function
(\ref{C2.17}) on ${\cal E}$.
Let $b_{*} \in {\cal E}$ be a point of maximum
of $f_{|}$.
Using the conditional extremum method
and the  fact that ${\cal E}$ is  ellipsoid
we  prove that $b_{*} = b_0$. (For brane vectors $U^s$ see \cite{IMp3}).

We consider the function
\begin{eqnarray} \label{C2.22}
 \bar f(b, \lambda) \equiv  f(b)
 - \lambda_0 U^0_i b^i  
 - \sum_{s \in S} \lambda_s U^s_i b^i
 \\ \nonumber
   - \lambda_2 \left(\sum_{s,s' \in S} \frac{h_s}{2} A_{s s'} b^s b^{s'}
   +  G_{ij} b^i b^j + h_0 \right),
\end{eqnarray}
where
$\lambda = (\lambda_0,  \lambda_s, \lambda_2)$ is a vector of
Lagrange multipliers. It can be readily verified that
the  points of extremum for the
function  $\bar f$ from (\ref{C2.22}) have the
form $(\lambda_2  b_{0}, \lambda)$ with $b_{0}$ defined by
relations (\ref{C2.19}) and (\ref{C2.20}), and $\lambda$ 
given by
\begin{eqnarray} \label{C2.23}
    \lambda_0 = 1/(d_0 -1), \quad
  \lambda_s = - 2 \sum_{s' \in S} h_s A^{s s'} U^{s'}_{1},
  \quad   \lambda_2 = \pm 1,
\end{eqnarray}
$s \in S$. Then, the  points $b_{*}$ and $ - b_{*}$
are the points of maximum
and minimum, respectively, for the function $f_{|}$ defined on the
ellipsoid  ${\cal E}$. Since $f(b_{0}) = |h_0|$, the only point
satisfying the restriction $f(b) \geq |h_0|$ is $b = b_{0}$.
From (\ref{C2.16}) we get
\begin{eqnarray} \label{C2.24}
\sum_{l \in S} A_{s l} b^{l}_0 = 2 U^s_{1} > 0
\end{eqnarray}
for all $s \in S$. The proposition is proved.

Now we introduce a new radial variable $R = R(u)$ by relations
\begin{eqnarray} \label{C2.28}
\exp( - 2\bar{\mu} u) = 1 - \frac{2\mu}{R^{d}} = F,
\qquad \bar{\mu} = \sqrt{C_0}, \quad
\mu = \bar{\mu}/ d >0,
\end{eqnarray}
$u > 0$, $R^{d} > 2\mu$ ($ d = d_0 -1$).
We put
\begin{equation} 
\bar{c}^i = 0, \quad  q^s(0) = 0. \label{C2.27f}
\end{equation}
$i= 0, \dots, n$; $s \in S$.
These relations guarantee the asymptotical flatness
(for $R \to +\infty$) of the $(2+d_0)$-dimensional section of the metric.

Let us denote
\beq{C2.28a}
H_s = f_s e^{- \bar{\mu} b^s_0 u },
\eeq
$s \in S$.

Then,  solutions (\ref{5.g}), (\ref{5.r}) may be written as follows
\begin{eqnarray} \label{C2.30}
 g= \Bigl(\prod_{s \in S} H_s^{2 h_s \sum_{i=0}^{n} U^s_i} \Bigr)^{1/(D-2)}
 \biggl\{ F^{-1} dR \otimes dR
 + R^2  d \Omega^2_{d_0}  \\ \nonumber
 -  \Bigl(\prod_{s \in S} H_s^{-2 U^s_1 h_s} \Bigr) F  dt \otimes dt
 + \sum_{i = 2}^{n} \Bigl(\prod_{s\in S}
  H_s^{-2 h_s U^s_i/d_i} \Bigr) h^{[i]}  \biggr\},
\end{eqnarray}

\begin{equation} \label{5r}
 \hat{\rho}^s = -  A_s F^{U^s_1 -1} R^{ - 2d_0} 
  \left[   \prod_{s \in S}  H_{s}^{- 2 \sum_{i=0}^n h_s U^s_i} \right]^{1/(D-2)} 
  \prod_{s' \in S} H_{s'}^{ - A_{s s'}} , 
  \end{equation}
 $s \in S$.

Here $A_s \neq 0$, $h_s =K_s^{-1}$; $K_s \neq 0$ and
the non-degenerate matrix $(A_{s s'})$ is defined by  relation (\ref{2.1A}).

Functions $H_s > 0$ obey the equations
\beq{C2.34}
 R^{d_0} \frac{d}{dR} \left( R^{d_0}
\frac{F}{H_s}   \frac{d H_s}{dR} \right) = B_s F^{U^s_1 -1}
\prod_{s' \in S}  H_{s'}^{- A_{s s'}},
\eeq
$s \in S$, where $B_s \neq 0$ are defined
in (\ref{B1.21}).
These equations follow from Toda-type equations (\ref{B1.20}) and
the definition   (\ref{C2.28}) and   (\ref{C2.28a}).

It follows from (\ref{C2.7}), (\ref{C2.13}) and  (\ref{C2.28a})
that there exist finite limits
\beq{C2.35a}
H_s  \to H_{s0} \neq 0,
\eeq
as $R^d \to 2\mu$, $s \in S$ ($ d = d_0 -1$).


From (\ref{C2.27f})  we get
\beq{C2.35} H_s (R = +\infty) = 1, \eeq
$s \in S$.

\end{document}